\documentclass[12pt]{article}

\usepackage{longtable}

\usepackage{scicite}

\usepackage{times}

\newenvironment{sciabstract}{%
\begin{quote} \bf}
{\end{quote}}

\newcounter{lastnote}

\title{PAMELA Measurements of Cosmic-ray Proton and 
Helium Spectra}

\author
{O. Adriani$^{1,2}$, G. C. Barbarino$^{3,4}$, G. A. Bazilevskaya$^{5}$, R. Bellotti$^{6,7}$,\\
M. Boezio$^{8}$, E. A. Bogomolov$^{9}$, L. Bonechi$^{1,2}$, M. Bongi$^{2}$, V. Bonvicini$^{8}$,\\
S. Borisov$^{10,11,12}$, S. Bottai$^{2}$, A. Bruno$^{6,7}$, F. Cafagna$^{7}$, D. Campana$^{4}$,\\
R. Carbone$^{4,11}$, P. Carlson$^{13}$, M. Casolino$^{10}$, G. Castellini$^{14}$, L. Consiglio$^{4}$,\\
M. P. De Pascale$^{10,11}$, C. De Santis$^{10,11}$, N. De Simone$^{10,11}$, V. Di Felice$^{10}$, \\
 A. M. Galper$^{12}$, W. Gillard$^{13}$, L. Grishantseva$^{12}$, G. Jerse$^{8,15}$,\\
A. V. Karelin$^{12}$, S. V. Koldashov$^{12}$, S. Y. Krutkov$^{9}$, A. N. Kvashnin$^{5}$,\\
 A. Leonov$^{12}$,
V. Malakhov$^{12}$, V. Malvezzi$^{10}$, L. Marcelli$^{10}$, \\ 
A. G. Mayorov$^{12}$, W. Menn$^{16}$, V. V. Mikhailov$^{12}$,
E. Mocchiutti$^{8}$, A. Monaco$^{6,7}$,\\  N. Mori$^{1,2}$, N. Nikonov$^{9,10,11}$, G. Osteria$^{4}$,\\
F. Palma$^{10,11}$, P. Papini$^{2}$, M. Pearce$^{13}$, P. Picozza$^{\oplus,10,11}$, C. Pizzolotto$^{8}$, M. Ricci$^{17}$,\\
S. B. Ricciarini$^{2}$, L. Rossetto$^{13}$, R. Sarkar$^{8}$, M. Simon$^{16}$,\\ 
 R. Sparvoli$^{10,11}$, P. Spillantini$^{1,2}$,
Y. I. Stozhkov$^{5}$, A. Vacchi$^{8}$, \\
E. Vannuccini$^{2}$, G. Vasilyev$^{9}$, S. A. Voronov$^{12}$,\\
 Y. T. Yurkin$^{12}$, J. Wu$^{13,*}$, G. Zampa$^{8}$, N. Zampa$^{8}$, \\ 
 V. G. Zverev$^{12}$
\\
\\
\normalsize{$^{1}$University of Florence, Department of Physics, I-50019 Sesto Fiorentino, Florence, Italy}\\
\normalsize{$^{2}$INFN, Sezione di Florence, I-50019 Sesto Fiorentino, Florence, Italy}\\
\normalsize{$^{3}$University of Naples ``Federico II'', Department of Physics, I-80126 Naples, Italy}\\
\normalsize{$^{4}$INFN, Sezione di Naples,  I-80126 Naples, Italy}\\
\normalsize{$^{5}$Lebedev Physical Institute, RU-119991, Moscow, Russia}\\
\normalsize{$^{6}$University of Bari, Department of Physics, I-70126 Bari, Italy}\\
\normalsize{$^{7}$INFN, Sezione di Bari, I-70126 Bari, Italy}\\
\normalsize{$^{8}$INFN, Sezione di Trieste, I-34149 Trieste, Italy}\\
\normalsize{$^{9}$Ioffe Physical Technical Institute,  RU-194021 St. Petersburg, Russia}\\
\normalsize{$^{10}$INFN, Sezione di Rome ``Tor Vergata'', I-00133 Rome, Italy}\\
\normalsize{$^{11}$University of Rome ``Tor Vergata'', Department of Physics,  I-00133 Rome, Italy}\\
\normalsize{$^{12}$Moscow Engineering and Physics Institute,  RU-11540 Moscow, Russia}\\
\normalsize{$^{13}$KTH, Department of Physics, and the Oskar Klein Centre for Cosmoparticle Physics,}\\
\normalsize{AlbaNova University Centre, SE-10691 Stockholm, Sweden}\\
\normalsize{$^{14}$IFAC, I-50019 Sesto Fiorentino, Florence, Italy}\\
\normalsize{$^{15}$University of Trieste, Department of Physics, I-34147 Trieste, Italy}\\
\normalsize{$^{16}$Universit\"{a}t Siegen, Department of Physics, D-57068 Siegen, Germany}\\
\normalsize{$^{17}$INFN, Laboratori Nazionali di Frascati, Via Enrico Fermi 40, I-00044 Frascati, Italy}\\
\normalsize{$^{*}$On leave from  School of Mathematics and Physics, China University of Geosciences,}\\
\normalsize{CN-430074 Wuhan, China}\\
\normalsize{$\oplus$ To whom correspondence should be addressed. E-mail: picozza@roma2.infn.it}
}

\date{}

\usepackage{graphics}
\usepackage{graphicx}
\usepackage{epsfig}

\usepackage{amssymb}

\usepackage[usenames]{color} 

\newcommand{\pam}{PAMELA}

\begin{document}


\baselineskip24pt


\maketitle


\begin{sciabstract}
Protons and helium nuclei are the most abundant components of the
cosmic radiation. Precise measurements of their fluxes are needed
to understand the acceleration and subsequent propagation of cosmic rays in 
the Galaxy. We report precision measurements of the proton and helium spectra in the
rigidity range 1 GV-1.2 TV performed by the satellite-borne experiment
PAMELA. We find that the spectral shapes of these two species are
different and cannot be well described by a single power law. These data
challenge the current paradigm
of cosmic-ray acceleration in supernova
remnants followed by diffusive propagation in the Galaxy. More complex
processes of acceleration and propagation of cosmic rays are required
to explain the spectral structures observed in our data. 
\end{sciabstract}
\maketitle

\section*{}

Since the discovery of cosmic rays, various mechanisms have been
proposed to explain the acceleration of particles to relativistic
energies and their subsequent propagation in the Galaxy.  
It was pointed out long ago [e.g. \cite{Legage1983,Ginzburg1964}] that
supernovae fulfill the power requirement to energize galactic
cosmic rays. Subsequently, models were put forward explaining the
acceleration 
of cosmic ray particles via diffusive shock acceleration 
produced by SN (supernova) shock waves propagating in the interstellar
medium [see \cite{2001RPPh...64..429M} for a review]. 

At the end of the acceleration phase, particles are injected into the
interstellar medium where 
they propagate, diffusing through the turbulent galactic magnetic
fields. Nowadays, this propagation  
is well described by solving numerically \cite{StrongMoskalenko1998}
or analytically \cite{2001ApJ...547..264J,Donato2001} the transport equations
for particle diffusion in the Galaxy. 
The galactic magnetic fields mask the arrival direction of charged
particles, making the cosmic-ray flux 
isotropic although there are hints of anisotropy in the 10-100 TeV
range \cite{amenomoriscience}.   

Recent \pam\ measurements of the antiparticle component of the cosmic
radiation\cite{naturepos,pbarratio,pbarflux} have prompted a
re-evaluation of possible contributions from additional galactic
sources, either of astrophysical [e.g. pulsars
\cite{2009CQGra..26w5009G}] or exotic [e.g. dark matter
\cite{2009PhRvD..79a5014A,2009PhLB..681..151K}] origin. 
Detailed knowledge of cosmic ray spectra is needed to: a)
identify sources and acceleration/propagation mechanisms of cosmic
rays; b) estimate the production of secondary particles, such as
positrons and antiprotons,  in order to disentangle the secondary
particle component from possible exotic sources; c) estimate the
particle flux in the geomagnetic field and in Earth's atmosphere for
in-orbit dose estimations and to derive the atmospheric muon and
neutrino flux, respectively. 

\begin{figure}[!ht]
\begin{center}
\includegraphics[width=25pc]{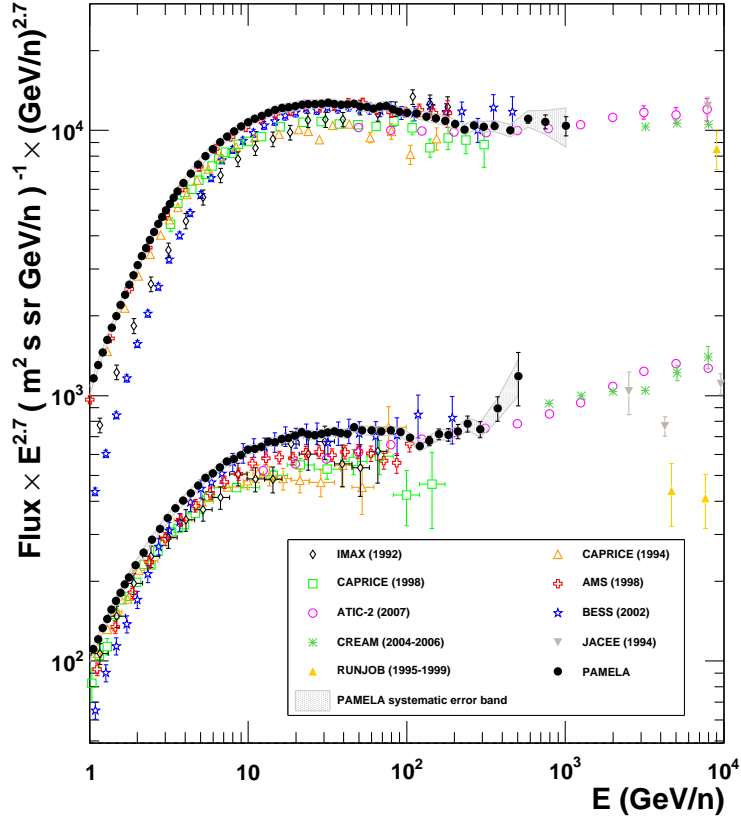}
\caption{Proton and helium absolute fluxes measured by PAMELA above 1
  GeV/n, compared with a few of the previous
  measurements~\cite{IMAX2000,CAP94_protons,CAP98_protons,AMS_protons,atic2icrc,besstev, creamdiscrepant,runjob,jacee1998}. 
All previous 
measurements but one \cite{AMS_protons} come from balloon-borne experiments.  
Previous data up to few hundred GeV/n  
were collected by magnetic spectrometer experiments
\cite{IMAX2000,CAP94_protons,CAP98_protons,AMS_protons,besstev}  
while higher energy data come from calorimetric measurements. 
PAMELA data
cover the energy range 1 GeV -1.2 TeV (1-600 GeV/n for He). 
The fluxes are expressed in terms of kinetic energy per 
nucleon, converted from the rigidity measured in the tracker and
neglecting any contribution from 
less abundant deuterium (d/p $ \simeq 1 \% $) and $^{3}He$
($^{3}He/^{4}He \simeq 10\%$).  
Pure proton and
$^{4}He$ samples are therefore assumed. 
Error bars are statistical, the shaded area represents the estimated
systematic uncertainty.}
\end{center}
\end{figure}

We present absolute cosmic ray proton and helium spectra
in the rigidity interval between 1 GV and 1.2 TV (Fig. 1, Tables S1,
S2), based on data gathered
between 2006-2008 with \pam, a detector orbiting the Earth in
a 350-610 km, 70$^\circ$ inclination orbit as part of the Russian
Resurs-DK1 spacecraft \cite{pamel}.

Our results are consistent with those of other
experiments (Fig. 1),  
considering the statistical and systematic uncertainties of the
various experiments. 
There are differences at low ($<30$ GeV) energies caused by  
solar modulation effects (\pam\ was operating during a period of
minimum solar activity with solar modulation parameter, $\phi$, of
450-550 MV  in the spherical force 
field approximation \cite{gleesonaxford1968}). 
PAMELA results overlap with ATIC-2 data \cite{atic2icrc}
between $\sim200$ GV and $ \sim1200$ GV, but  differ both in shape and
absolute normalisation at lower 
energies. The extrapolation to higher energy of the PAMELA fluxes
suggest a broad agreement with those published by 
CREAM \cite{creamdiscrepant} and JACEE \cite{jacee1998} but are higher
than the RUNJOB \cite{runjob} helium data.  

\begin{figure}[!ht]
\begin{center}
\includegraphics[width=25pc]{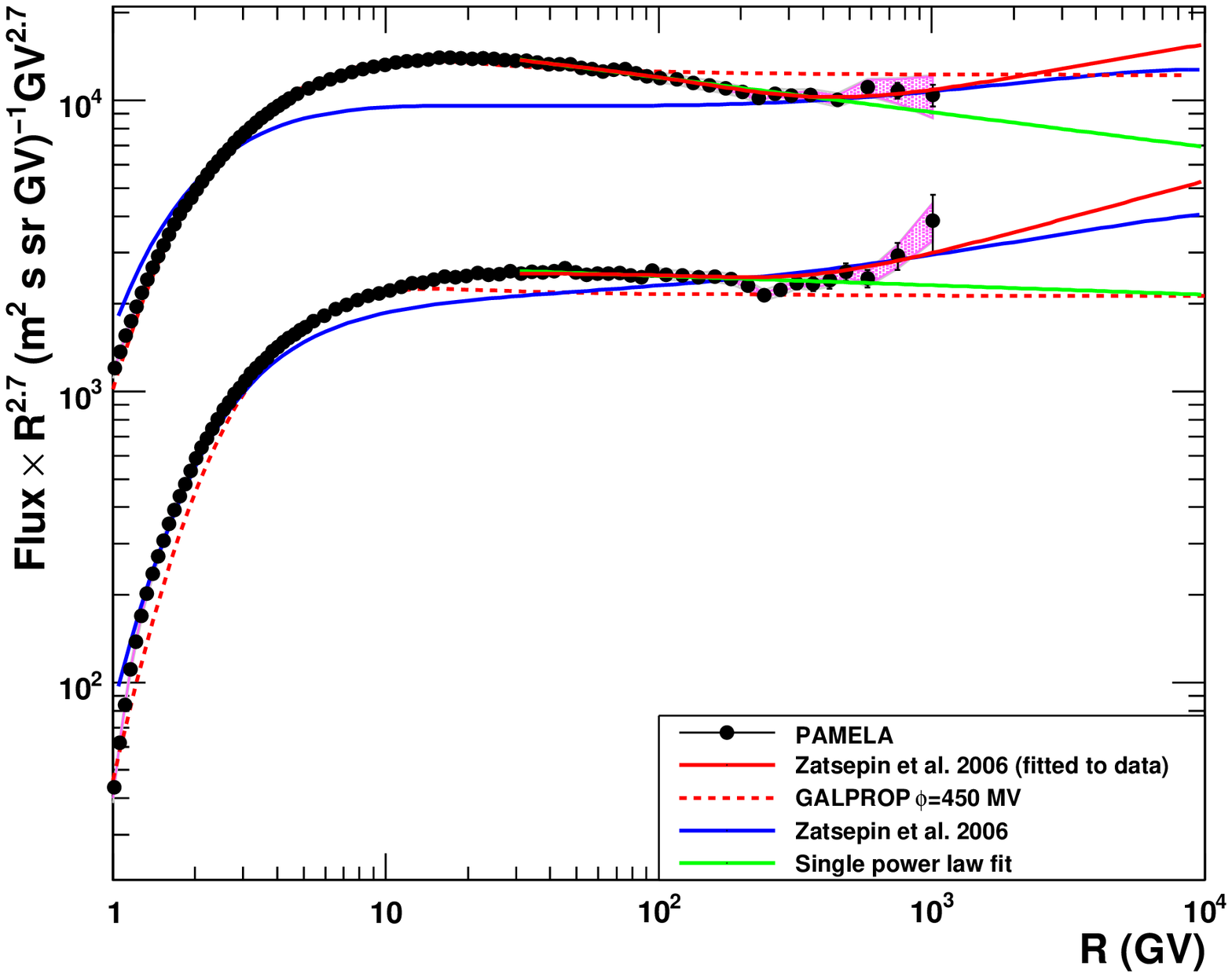}
\caption{Proton (top points) and helium (bottom points) data
   measured by \pam\ in the rigidity range 1 GV - 1.2 TV. The shaded
   area represents the estimated systematic uncertainty. The lines
   represent the fit with a single power law and the
   Galprop\cite{webgalprop} and
   Zatsepin\cite{zatsepin2006} models. Details of the models are
   presented in Tables S1, S2. }
\end{center}
\end{figure}
To gain a better understanding of the spectra, we have analysed
our results in terms of rigidity instead of kinetic
energy per nucleon (Fig. 2 and Tables S3, S4). 
Two important conclusions can be drawn from the \pam\ data. 

Firstly,
the  proton and 
helium spectra ($J(R)$) 
have different spectral shapes. 
If a single power law, $J(R) = A
R^{-\gamma^{R}}$, is fit to the data 
between 30 GV (above the influence of solar modulation)
and 1.2 TV, the resulting spectral indices are:
\[
\gamma^{R}_{30-1000 \: {\rm GV,p}} = 2.820\pm 0.003 (stat) \pm 0.005 (syst), \]
\[
\gamma^{R}_{30-1000\: {\rm GV,He}} = 2.732\pm 0.005 ({stat}) ^{+0.008}_{-0.003} 
(syst), 
\label{fittone}
\]
which establishes that there is a significant difference between the
two spectral indices in this rigidity region.
\begin{figure}[!ht]
\begin{center}
\includegraphics[width=25pc]{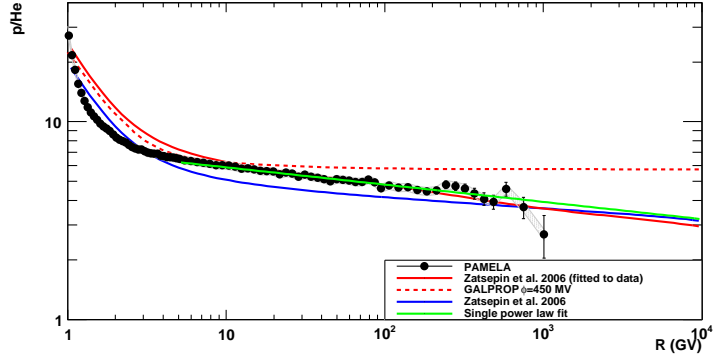}
\caption{Ratio of the flux between proton and helium
data of \pam\ 
   vs.  Rigidity. The shaded area represents the estimated systematic
   uncertainty. Lines show the fit using one single power law
   (describing the difference of the two spectral indices), the 
   Galprop\cite{webgalprop} and Zatsepin  models 
   with the original values of the paper\cite{zatsepin2006} and fitted
   to the data.  Details of the models are
   presented in Tables S1, S2.}
\end{center}
\end{figure}
These effects are also seen in Fig. 3 (and in Table S5), where the
proton-to-helium flux ratio is shown as a 
function of rigidity. Presenting the results as a ratio reduces the
possible impact of systematic errors because  
a number of instrumental effects cancel in the ratio, e.g. the
estimation of live time and the error  
associated with the alignment of the tracker and the track
reconstruction algorithm. 
The proton-to-helium flux ratio shows a continuous and smooth decrease
as the rigidity increases. The same ratio cast in terms of kinetic
energy per 
nucleon or total kinetic energy exhibits more irregular behaviour
(Fig. S1). 
By applying a power law approximation to the two spectra, the ratio can
be used to determine the difference 
between the two spectral indices with a smaller associated systematic error, 
$\Delta_{\gamma^{R}}=\gamma^{R}_{p}-\gamma^{R}_{He} = 0.101 \pm 0.0014
(stat) \pm 0.0001 (sys)$.
The ratio is well described
by a power law down to rigidities as low 
as 5~GV (green line in Fig. 3).   
For rigidities $R>>\phi$, the ratio of the two species is
independent of the solar modulation parameter and allows 
$\Delta_{\gamma}$ for the interstellar spectrum to be measured in the rigidity range
5-30 GV, where solar modulation effects dominate.   
Previous measurements
\cite{IMAX2000,CAP94_protons,CAP98_protons,besstev,AMS_protons} did
not have the statistical and systematic precision to demonstrate this
decrease in the ratio. 

\begin{figure}[!ht]
\begin{center}
\includegraphics[width=25pc]{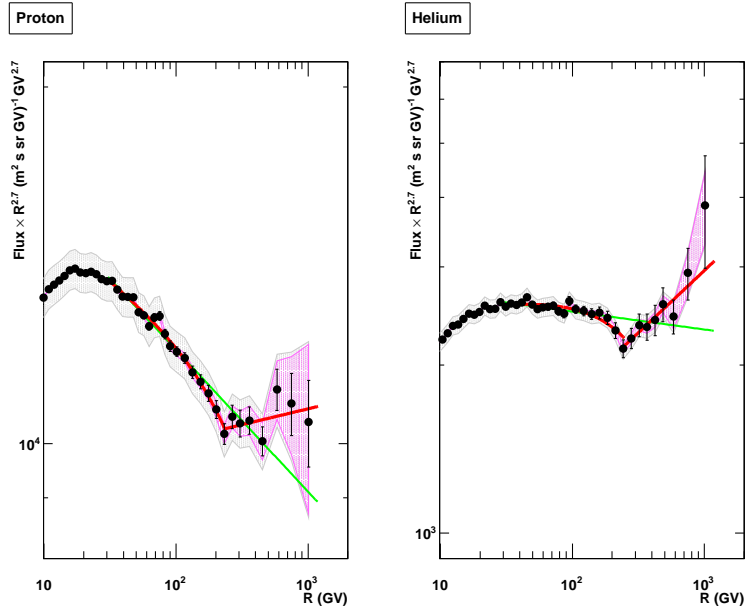}
\caption{Proton (left panel) and helium (right panel)
spectra in the range 10 GV - 1.2 TV.  
 The grey shaded area represents the estimated systematic uncertainty,
 the pink shaded area represents the contribution due to tracker
 alignment. The straight (green) lines represent fits with a single
 power law in the rigidity range 30 GV - 240 GV. The red curves
 represent the fit with a rigidity dependent power law (30-240 GV) and
 with a single power law above 240 GV. }
\end{center}
\end{figure}
Secondly, as seen in Fig. 4, the PAMELA 
data show clear deviations from a single power
law model: 

The spectrum of protons gradually softens in the rigidity range
30-230 GV. In the rigidity range 30-80 GV,  
$\gamma^R_{30-80\: {\rm GV, p}} = 2.801 \pm 0.007 \: (stat)  \pm 0.002 \:
(syst)$, which is lower than the value fitted
between 80-230 GV:  
$\gamma^R_{80-230\: {\rm GV,p}} = 2.850 \pm 0.015 (stat) \pm 0.004 (syst)$. 
In the case of helium, $\gamma^R_{30-80\: {\rm GV, He}} = 2.71 \pm 0.01 \:
(stat)  \pm 0.002 \: (syst)$, which is 
lower than $\gamma^R_{80-230\: {\rm GV,He}} = 2.77 \pm 0.03 (stat) \pm
0.004 (syst)$.  
We applied Fisher's  and Student's $t$-tests 
to the single power law hypothesis in the range 30-230 GV for both
protons and helium (see Section 5 of the Supporting Online
Material (SOM\cite{som}) for details). 
This hypothesis is rejected at 95\% confidence level (C.L.).
Considering the same rigidity interval in terms of kinetic energy per
nucleon 
the Fisher's and Student's $t$-tests
reject a single power law hypothesis at 99.7\%  C.L.. 

At 230-240 GV the proton and helium data exhibit an abrupt spectral
hardening.  
Applying Fisher's test and Student's $t$-test  to the proton spectrum
above 80 GV, the single power law hypothesis is rejected at  99.7\%
C.L. if only statistical errors are considered.  
A similar result is obtained if the fluxes are increased in line with
the systematic uncertainties.  
If the fluxes are instead decreased, the single power law hypothesis
is rejected at 95\%  C.L..  
The hardening of the proton spectrum occurs at $232^{+35}_{-30}$ GV
with change of spectral index from  
$\gamma^{R}_{80-232 {\rm GV,p}}=2.85\pm 0.015 \pm 0.004 $ to 
$\gamma^{R}_{>232 {\rm GV,p}}=2.67\pm 0.03 \pm 0.05 $. 
For the helium data, the single power law hypothesis is rejected at
95\% C.L. with spectral hardening setting in at  
$243^{+27}_{-31}$ GV and a corresponding change of spectral index of
$\gamma^{R}_{80-240 {\rm GV,He}}=2.766\pm 0.01 \pm 0.027 $  
and $\gamma^{R}_{>243 {\rm GV,He}}=2.477\pm 0.06\pm 0.03$.  
As a consistency check, we repeated this analysis 
with the three highest energy data points excluded: no changes in the
proton and helium results were observed. 
We obtained similar results when using
alternative statistical methods such as the cumulative sum
test (see Section 5.4 in the SOM\cite{som}). 

One of the most striking features of the cosmic rays prior
to \pam\ observations was 
their apparently featureless energy spectra. Until now, single power
laws, as predicted by the shock diffusion acceleration model and
diffusive propagation in the Galaxy [see \cite{2007ARNPS..57..285S}
for a recent review], could reproduce spectra using similar
spectral indices (a fit to the experimental data yields $\gamma \simeq
2.7$) for protons and heavier nuclei up to energies of 
about $\approx 10^{15}$ eV (the so-called `knee' region). Such 
assumptions are routinely incorporated into common used
propagation models, such as GALPROP
\cite{StrongMoskalenko1998}, which is widely considered to be the 
standard model of cosmic-ray acceleration and propagation. 
Our results challenge this scenario \cite{solar}. 
As it can be seen in Figs. 2 and 3 
the GALPROP calculation does
not reproduce PAMELA data across the full rigidity region. 
Moreover it is difficult, even with recent models of
non-linear shock acceleration [e.g. \cite{cap10,2007ApJ...661..879E}],
to 
produce significant differences in the proton and helium spectra as
low as a few tens of GV.   

The hardening in the spectra observed by PAMELA around 200 GV could
be interpreted as an 
indication of different populations of cosmic ray
sources.  
As an example of a multi-source model, Fig. 2 shows a
comparison  
with a calculation (blue curves) by Zatsepin and
Sokolskaya~\cite{zatsepin2006}, which was put forward to explain
ATIC-2 data \cite{atic2icrc} and 
considered novae stars and 
explosions in superbubbles as additional cosmic-ray sources. 
The parameters of the model were fitted to
match  ATIC-2 data and, consequently, are in disagreement with 
PAMELA data
in  absolute fluxes and the ratio.  If the parameters of this model
are fitted to the \pam\ data the agreement can be greatly improved
(red curves of Fig. 2 and 3).  
CREAM also reported a direct measurement, albeit with a low
statistical and systematic significance, of a change of the slope for
nuclei (Z $\geq$ 3) at 200 GeV/n, i.e. at a higher rigidity ($\simeq
400$ GV) than our observed break in helium spectrum. 

An indication that proton and helium have different spectral indices
at high energy ($\sim 10$ TeV) was reported by
JACEE~\cite{jacee1998}. More recently CREAM~\cite{creamdiscrepant}
indirectly inferred [using also AMS~\cite{AMS_protons} and
BESS~\cite{BESS_p} data)] that spectral deformation should 
occur at about 200 GeV/n for both species. This is similar to
our results for protons but higher (400 GV) than our results for
helium. 
Results from ATIC-2~\cite{atic2icrc} implied that protons and helium
nuclei have different energy spectra,  
although the results suffered from 
unclear systematic uncertainties  and there were differences with
respect to previously reported ATIC-1~\cite{Ahn2006} data.

\bibliographystyle{Science}

\bibliography{1199172Revisedtext}

\begin{thebibliography}{10}
\bibitem{pamel}
{PAMELA comprises a number of high performance detectors, capable of
  identifying particles through the determination of charge (Z), rigidity
  ($R=pc/|Z|e$, p being the momentum of a particle of charge Ze) and velocity
  ($\beta=v/c $) over a wide energy range. The device is built around a
  permanent magnet with a 6 plane double-sided silicon micro-strip tracker,
  providing absolute charge information and track-deflection ($ \eta = \pm
  1/R$, with the sign depending on the sign of the charge derived from the
  curvature direction) information. A scintillator system, composed of three
  double layers of scintillators (S1, S2, S3) provides the trigger, a
  time-of-flight measurement and an additional estimation of absolute charge. A
  silicon-tungsten tracking calorimeter, a bottom scintillator (S4) and a
  neutron detector are used to perform lepton-hadron discrimination. An
  anticoincidence system is used off-line to reject spurious event triggers
  generated by particles interacting in the apparatus. Respect to balloon-borne
  experiments, PAMELA has the advantage of a significantly longer period of
  uninterrupted observing time. Furthemore, data taking in space is not
  affected by enviromental systematics such as those due to correction for
  secondary particles produced in the residual atmosphere that affects
  balloon-borne experiments. A more detailed description of PAMELA and the
  analysis methodology can be found in~\cite{Pi07,launchofpamela} and in the
  Supporting Online Material.}
\end{thebibliography}


\clearpage
 \nocite{Pi07,launchofpamela,jokipii1977,potgieter2008,thankssss}

\end{document}